# Scheme of Derivation of Collapse from Quantum Dynamics

by Roland Omnès

Laboratoire de Physique Théorique (Unité Mixte de Recherche, CNRS, UMR 8627), Université de Paris XI, Bâtiment 210, 91405 Orsay, Cedex, France
e-mail: Roland.Omnes@th.u-psud.fr

———————

*Abstract*

Two categories of results regarding quantum measurements are derived in this work and applied to the problem of collapse.

The first category is concerned with local and transient features of the entanglement between a macroscopic measuring system and a measured one. These properties result directly from the Schrödinger equation. They cannot be formulated in terms of observables, do not affect the wave functions themselves but express their history in an irreversible way. They carry a specific new kind of local probabilities, which evolve with a finite velocity under nonlinear wave equations.

The second category of results extends these local properties to the case of a macroscopic system and its environment. Fluctuations in their interaction are predicted then and generate a specific incoherence in the quantum state of the system.

These two kinds of effects act together when a macroscopic measuring system interacts with a measured system and with an environment. Their combination yields then an explicit and effective mechanism of collapse, with a random behavior resulting from random incoherence. Born's basic probability rule for the results of measurements turns out then simply a consequence of quantum dynamics.

Some conjectures still enter into the derivation of these effects, which one may recognize to look hardly credible at first sight. They fit however so well together that one proposes a more thorough investigation of their approach as a promising strategy for a self-contained explanation of collapse.

———————

The problem of wave function collapse —asking why reality is unique in a quantum world— came to the foreground of physics long ago and remains a nagging question: "Can one really understand quantum mechanics?" A series of articles by Schrödinger [1] and one by Einstein, Podolsky and Rosen [2], both published in 1935, left a long-lasting impression of an incompleteness in this regard [3,4].

Many answers were attempted. Some essays tried to modify in depth the interpretation of quantum mechanics [5-7], or revised its foundations [8, 9]. Efforts were made also to complete its physical content by extraneous phenomena [10], or questioned its exactness [11, 12]. Other conceptions of understanding were proposed along more philosophical lines [13-15].

The approach in the present work is rather more conventional since its aim is to propose (on the basis of some new results) that *quantum mechanics could predict collapse as a consequence of the Schrödinger equation*.

A first hint along that direction came from local properties of entanglement, which were discovered by Lieb and Robinson, in the framework of spin lattices [16]. They were rediscovered by the present authors and extended recently with new suggestions [17]. These local properties (which exist only in a macroscopic system) derive directly from the Schrödinger equation, although they cannot be expressed in terms of observables. They evolve nevertheless. When a charged particle enters a wire chamber, for instance, local entanglement begins in a restricted region, where the particle acts directly on atoms in the chamber. Soon after, this local entanglement grows, during a short but sizable time, until it extends to the whole chamber. Standard algebraic entanglement with the charged particle is then completely reached.

These local properties of entanglement play a central part in the present work and their behavior is often surprising. There exists for instance a probability of local entanglement $f_1(x, t)$, at a time $t$ near any point $x$ in the chamber, and also a probability $f_0(x, t)$ for the absence of local entanglement. These probabilities are unrelated with standard quantum ones (since they are unrelated with observables). Their time dependence shows a macroscopic evolution of local entanglement with remarkable features, since it is *irreversible* and cannot be spelled out by an observable on a wave function, at any specific time: Only the previous *history* of evolution of this function, till that time, can show that this effect exists.

Another consequence of local entanglement was found during the present work. It consists in *a generation of incoherence* in the state of a macroscopic system, by the fluctuations in the influence that this system receives from its environment. It will be described here, but one must recognize that some of its features are still conjectural. Their full understanding would need a systematic investigation, which would be certainly difficult. The possibilities that it raises, nevertheless, look so promising that one will sketch them here as indicating a new way of research with attractive prospects.

The problem under study is thus concerned with a macroscopic quantum system (a measuring device) and the relation with its environment. This situation has many variants but one will consider only a case where the environment is an external atmosphere. The results regarding local entanglement can be applied in that case to the interaction of a unique atmospheric molecule, when it hits the apparatus, but there are many molecules in the environment and these results must be extended. The problem of this extension obstructed this research for a long time, until a simple proposal came out at last, as follows: The environment has no remarkable quantum effect by itself on average (no more than a thermostat does for instance). *Fluctuations* in the action of environment can be essential on the contrary regarding quantum details. Their main consequence is a permanent generation of some incoherence and also much disorder into the quantum state of a macroscopic system.

This result is proposed as a key, by means of which an explicit mechanism for the occurrence of collapse would derive from the rules of quantum dynamics, under an influence of environment.

In the simple case of a measuring device containing an atomic gas, one can sketch this mechanism as follows: The initial state of the measured system is contains different measurement channels, corresponding with different values of a measured observable. Let one label these channel by an index $j$. The existence of collapse would follow then from an elementary effect, in which for instance two atoms in the gas inside a wire chamber collide. The state of the first atom, denoted by *a*, is locally entangled with some measurement channel *j*. The state of the second atom, denoted by *b*, has no such local entanglement with any channel *and* is moreover incoherent (as a consequence of incoherence from external fluctuations). The collision of the two atoms in these specific states yields then a "*slip in coherence*", in which small fractions of the quantum probabilities of measurement channels with indices $j' \neq j$ are transferred to the channel *j*!

The existence of this elementary effect, under which small random exchanges occur between the quantum probabilities of different channels, is the main result of the present work.

The resulting theory is developed as follows in this paper. Section 1 indicates a few necessary reminders regarding quantum foundations [18,19], including particularly the principle of cluster decomposition [20], rarely mentioned if ever in measurement theory but closely linked with local entanglement.

Section 2 introduces this local entanglement, simplified to get more rapidly at its main features regarding collapse. Section 3 is more technical and provides a mathematical derivation of local entanglement from the Schrödinger equation, with discussion of its main properties.

Section 4 deals with the action of an environment (which must play a part in an actual measurement, since one knows that collapse would never occur otherwise [1, 21]). One finds that *fluctuations* in the action of environment are responsible for a significant amount of incoherence in the quantum state of a macroscopic measuring system, as a consequence of its local entanglement with elementary constituents of the environment (for instance molecules in a surrounding atmosphere). These results are new and show striking features. The main one is that neither the direct action of environment nor even the fluctuations in this action can be direct agents of collapse, but a competition between positive and negative fluctuations is probably responsible for this effect.

Section 5 describes an explicit theory and a mechanism of collapse. After explaining slips in coherence, it shows how their accumulation yields random shifts among the quantum probabilities of various measurement channels. The outcome is a collapse phenomenon, ending with the emergence of a unique measurement datum. The underlying randomness of this collapse process holds in the fluctuations in action of the environment on the measuring device. Born's probability rule comes out under these conditions from a theorem by Philip Pearle [11]. Only one point in the proof would still need improvement and maintains a part of its tantalizing attraction in the charm of the problem.

Section 6 consists only in several comments, illustrations and further questions regarding the results.

The paper is not short, because many aspects of measurement theory need reconsideration along the way, but one tried to limit its length.

**1. Preliminaries**

The present preliminaries involve a combination of experimental, theoretical and conceptual considerations. They intend to provide a convenient frame for research on the topic of collapse according to several points or assumptions as follow.

1. The basic rules of quantum mechanics, regarding wave functions, observables and dynamical laws under the Schrödinger equation, will be considered universal and expressed in the Dirac- Von Neumann formalism [18, 19].

2. The purpose of this work is not to question the intrinsic randomness of quantum processes or the associated quantum probabilities, expressed by squares of quantum amplitudes according to Born's probability rule. One will take them as granted here. The problem in which one is interested consists then in showing that collapse is a consequence of the quantum laws, and in understanding better why it occurs randomly and why its proper macroscopic randomness coincides with Born's rule, which expresses an intrinsic microscopic randomness.

3. Macroscopic dynamics, which uses the Lagrange-Hamilton formalism for collective observables at macroscopic scales, is considered a consequence of Schrödinger's dynamics at these scales, according to previous derivations of classical physics [22]. The conservation of uniqueness of this classical world at macroscopic scales is therefore insured, but the origin and emergence of this uniqueness raise difficult questions, which belong to the topics of the present work.

4. Careful experiments, especially in quantum optics, have shown decisively that collapse does never occur at a microscopic level [23]. It always happens on the contrary in measurements at macroscopic scales, so that measurement theory can only be concerned with macroscopic measuring devices.

5. Local properties of entanglement take a central place in the present approach. They were discovered by Lieb and Robinson in the framework of spin lattices [16], but never came to the center of attention in research regarding collapse. The present author, who rediscovered them in that framework, called them "*intricacy*", in ignorance of their previous existence [17], but one will keep its initial name of local entanglement, in spite of some risk of confusion when local entanglement is opposed to global (algebraic) entanglement [24], as often happens when problems regarding measurements are considered.

6. The existence of local entanglement illustrates a principle of "cluster decomposition" in quantum theory, which is not often mentioned among the principles of the theory (at least in the literature on interpretation). Steven Weinberg stressed the necessity of this principle for a consistent axiomatic construction of quantum field theory, not starting as usual from a quantized version of classical electrodynamics ([20], I, Chapter 4).
Wichmann and Crichton gave a first explicit expression of this principle [25], which "says in effect, that distant experiments yield uncorrelated results" [20]. The idea had however appeared previously in statistical physics, as clustering in Green functions, partition functions or resolvents [26-28]. It occurred also in studies of multiple scattering and particularly Faddeev's equations [29]. The cluster decomposition principle has therefore many faces, including derivations of Feynman paths in field theory and of Faddeev-Popov ghosts in gauge theories ([20], II).

Many forms of "clusters" express a decrease of algebraic connection with distance and, in that sense, a weakening in the constraints arising from the framework of Hilbert spaces and observables. No complete and rigorous mathematical construction is known however for them (as far as the present author is aware) but one will see other examples of their properties in the case of collapse.

7. Another significant point is concerned with the status of probabilities in quantum mechanics. Von Neumann considered that every probability belonging to quantum theory should necessarily refer to a projection operator expressing a value of an observable [18]. In a plainer language, every probability is the square of some quantum amplitude.

Although this definition is perfectly proper, one will consider nevertheless that its domain does not extend as far as excluding any meaning for other kinds of probabilities, which would be useful and significant in the *interpretation* of quantum mechanics.

This standpoint is not new and an earlier example occurred with consistent histories [30-32], which involved a wider notion of probability, specific to histories of a system. They got benefit from it in logical matters, by discarding paradoxes that had for a long time obscured interpretation [33]. Other types of probabilities, involving local entanglement, will be found especially useful here.

8. A nontrivial question, which will also become of interest, is concerned with the notion of quantum system as compared with the notion of environment, especially regarding the question: Can one consider the environment of a quantum system as being itself a quantum system?

The main point is that a quantum system, say $S$, is associated with a well-defined set of observables (a C*-algebra), which defines all its properties. It is also associated with a state (expressed by a density matrix $\rho_S$), which yields quantum probabilities for these properties. Fulfillment of these two conditions is not obvious in the case of the environment around a system $S$.

There is a vast literature on this question where, briefly said, a party holds for the existence of a wave function $\Psi_U$ of the universe [7] and the opposition rejects this existence as meaningless. One will take here a middle way where $\Psi_U$ will not be given a grand meaning, but will only be used to define the state $\rho_S$ of a system $S$ as a partial trace over the product $|\Psi_U\rangle\langle\Psi_U|$. One will avoid however using $\Psi_U$ as a source of quantum correlations between local objects and the total universe, with ultimate branching effects extending from an event in a local apparatus to this totality. The use of $\Psi_U$ will mean therefore only that the same quantum laws hold for every real system in which one is interested, as they do at every place in its environment and everywhere else.

9. Another preliminary is concerned with a theorem by Philip Pearle, which is central in CSL theories [11, 12] and provides a remarkable framework for understanding what could be the origin collapse.

This theorem assumes that the initial state of a quantum system is a superposition

$$|A\rangle = \Sigma_j c_j |j\rangle, \tag{1.1}$$

of state vectors $|j\rangle$, which are eigenvectors of a measured observable. The quantum probabilities $p_j$ for the various channels start from the values $p_j = |c_j|^2$ and are assumed to undergo random motion in an approach to collapse: The fluctuations $\delta p_j$ during a short time intervals $\delta t$ have then vanishing average values and correlations

$$< \delta p_j \delta p_k > = A_{jk} \, \delta t, \qquad\qquad\qquad (1.2)$$

where the linear behavior in $\delta t$ is characteristic of a random process and the coefficients $A_{jk}$ are supposed to depend only on time and on the quantities $\{p_j\}$. Under a few more conditions, which will be examined in Section 5, Pearle's theorem asserts that a unique quantum probability $p_j$ must finally reach the value 1 after that other ones have successively vanished. Moreover –and this is the crux of the matter– the Brownian probability for this event is equal to $\mid c_j \mid^2$ so that Born's probability rule becomes a theorem. The existence of this theorem was among the main guides in the present work.

## 2. Local entanglement

Local entanglement is based on some mathematical properties of the Schrödinger equation, which will be explained in more detail in the next section. The present one is only introductive.

It begins like an exercise about Schrödinger's equation: One considers a particle, denoted by $A$. A gas of atoms, enclosed in a box, constitutes another quantum system $B$, which is macroscopic. The atoms in $B$ interact together through two-body interactions under a potential $V$. They can also interact with the particle $A$ under another potential $U$. The particle $A$ arrives initially from outside and enters the box at time zero.

One knows then how to write down the Schrödinger equation governing the composite system $AB$. But one can also play a sort of game with the intention of following in some detail how the particle influences gradually the atoms in different regions of the box containing the gas.

Let one assume for clearer illustration that the particle $A$ carries a red color and all the atoms are initially colored white before entry of the particle into the box. The rules of the game of influence specify that when the red particle $A$ interacts with a white atom $a$, this atom becomes red. When such a red atom $a$ interacts later with another atom $b$, which is still white, $b$ becomes red and $a$ remains red. The red color is acquired moreover once and for all by an atom so that, when two red atoms interact, they remain red, and two white atoms remain similarly white when they interact. Because of this conservation of the red color, when the particle $A$ interacts with an already red atom, this atom remains red. The rules of the game are then complete and the acquisition of redness is an irreversible process, which one will call a "contagion" with an obvious analogy. There is another analogy of the white and red colors with a two-valued quantum number.

All of this is detailed in the next section as an exercise using the Schrödinger equation. The white color of an atom is then associated with a value 0 for a kind of quantum number and the red color with the value 1 for that number. One will also say that an atom in a state with index 1 is *locally entangled* with the particle $A$ and is non-locally entangled in the case of index 0.

This evolution of local entanglement can be expressed in quantum terms by means of a Schrödinger equation in which atoms are supposed to carry a quantum number of local entanglement (or a color keeping memory). This quantum behavior can also be compared with the case of a classical motion of the atoms, in the case of short-range collisions. Remarkable properties of local entanglement appear then, at which one will now look.

*Probabilities of local entanglement and of non-local entanglement*

When the atoms and the particle $A$ move classically and are assimilated with hard spheres, there must exist, near any point $x$ in the gas and at any time $t$, a probability of local entanglement $f_1(x,t)$ and a probability of non-local entanglement, $f_1(x,t)$, which satisfy the relation

$$f_1(x,\ t) + f_0(x,t) = 1 \qquad (2.1)$$

In the next section dealing with the quantum background of local entanglement, one will also consider a case where the state of the incoming particle $A$ is an eigenvector $|\,j\,\rangle$ of an observable before a measurement, or a linear combination as in Equation (1.1). One will find then as many local probabilities of local entanglement $f_j(x,t)$ as the number of measurement channels, together with a unique probability for non-local entanglement, given by the expression

$$f_0(x,\ t) = 1 - \Sigma_j\, p_j f_j(x,t) \qquad (2.2)$$

One needs not deal further presently with quantum aspects and one will now look at their macroscopic manifestation.

*Waves of local entanglement*

The question to considered here is concerned with the evolution of a probability of local entanglement $f_j(x,t)$. This is in principle a consequence of the Schrödinger equation, which will be shown in the next section but, presently, one wants only to illustrate its macroscopic manifestation in the case of a unique channel in which particle $A$ crosses the box containing the gas along a linear track. There are then only two local probabilities, $f_1(x,\ t)$ for local entanglement and $f_0(x,\ t)$ for non-local entanglement, which sum up to 1 according to Equation (2.1).

Simple considerations predict the evolution of $f_1(x,\ t)$. They rely on standard approximations in transport processes (such as heat diffusion or electric conduction for instance) where one considers that, in spite of the complexity of quantum processes, a transport process depending only on collisions between atoms can be approximated by a diffusion process [34]. The transport of local entanglement under two-atoms collisions can be then described by a diffusion equation

$$\partial f_1/\partial t_{\text{ diffusion}} = D\Delta f_1, \qquad (2.3)$$

where $D$ is a diffusion coefficient.

A change in the state of local entanglement occurs however, in addition to diffusion, when an atom, which is non-locally entangled with probability $f_0$, collides with a locally entangled atom. The first one becomes then locally entangled, by contagion of local entanglement, while the second one remains locally entangled as it was beforehand. The probability for this collision to occur near a point $x$ in the box during a short time interval $\delta t$ is then the product $f_1(x,\ t)\,f_0(x,\ t)\ \delta t/\tau$, where $\tau$ is the mean free time between successive collisions of an atom. The increase in $f_1(x,\ t)$ owing to contagion is therefore

$$\partial f_1/\partial t \text{ contagion} = f_1 f_0/ \tau. \qquad (2.4)$$

In spite of the approximate character of this expression, one will consider it as sufficiently valuable for representing the gross features of interesting effects. One gets accordingly, using (2.1) into account, a nonlinear diffusion equation

$$\partial f_1/\partial t = D\Delta f_1 + f_1(1 - f_1)/\tau. \qquad (2.5)$$

Simple considerations, which need not be developed here, show that Equation (2.5) cannot be satisfied by an everywhere positive function $f_1(x, t)$, contrary to the diffusion equation (2.3). The nonlinear character of Equation (2.3) requires therefore a domain of definition, or the existence of a finite region in space, bounded by some moving two-dimensional surface $S$ and containing the track of Particle $A$. In such a domain, $f_1(x, t)$ is positive, and it vanishes beyond the moving boundary. This behavior is frequent and moving wave fronts are often consequences of nonlinear wave equations, of which (2.5) is an example [35].

The wave front is essentially cylindrical around the track of Particle $A$ in the present example, when this particle is heavy and rapid enough. The probability of local entanglement $f_1(x, t)$ is still zero (indicating complete non-local entanglement) in the region besides the front, where the influence of $A$ has not yet been felt. Behind the front, the distribution $f_1(x)$ is shown in Figure 1, which is obtained from a numerical calculation in one space dimension.

One can expect essentially the same this behavior for a wave of local entanglement dimension 1 (in the ideal case of an excitation along a plane), dimension 2 (the cylindrical case) and 3 (for a point source). One notices also that the probability of local entanglement $f_1(x)$ becomes very close to 1 at a distance larger than a mean free path $\lambda$ behind the front, as shown in Figure 1.

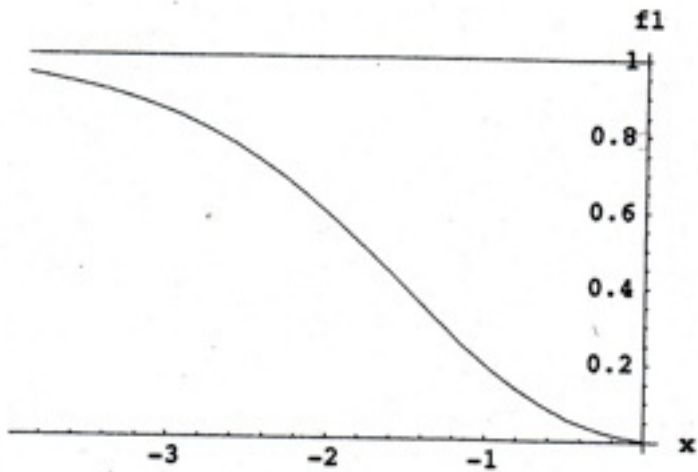

Figure 1: A graph of a wave of local entanglement $f_1(x)$ near its front (located here at $x = 0$), in one dimension or in the case of a cylindrical wave in three dimensions. The scale of abscissas is the mean free path $\lambda$ and the diffusion coefficient is taken as $D = \lambda^2/6\tau$.

One may notice also the value of the expansion velocity of a wave of local entanglement. The calculation leading to Figure 1 was made with use of a value $D = \lambda^2/6\tau$ for the diffusion coefficient, a relation resulting froù a random walk of atoms in three dimensions. The wave moves then at the velocity $v' = v/3^{1/2}$, which is the average velocity of an atom along the direction normal to the front. The point of interest here is that that this velocity coincides with the velocity of sound in a dilute gas. Whether this relation is general or not remains an unsolved problem though it seems sensible from the standpoint of fluid mechanics [36].

There are other interesting cases. A possible transmission of local entanglement by an electric current, under interactions of conducting electrons with impurities, suggests an associated velocity equal to the Fermi velocity in the conductor. Similarly, in a Geiger counter, a rapidly moving charged particle $A$ can excite or ionize nearby atoms. Photons from the decay of excited atoms would also transport local entanglement away at the velocity of light (though with retardation effects owing to the finite lifetime of excited states). Electrons, produced together with ions, would also transport local entanglement in their own way. This process of local entanglement appears therefore remarkably rich and strongly mingled with many other irreversible effects.

## 3. Quantum theory of local entanglement

This section is devoted to a mathematical theory of local entanglement using the Schrödinger equation, with examination of some of its properties, especially its irreversible behavior.

*Dynamics of local entanglement*

One uses again the model where a macroscopic system $B$ is a gas of atoms interacting through a two-body potential $V$. Another system $A$ consists in a particle, which can interact with every atom through another two-body potential $U$. Local entanglement is again associated with an index 1 and non-local entanglement with index 0.

A wave function $\psi$ of the $AB$ system evolves under the Schrödinger equation

$$i\hbar \partial \psi / \partial t = H\psi . \qquad (3.1)$$

One can again take account of the influence of Particle $A$ on the gas by considering that every atom carries an index of local entanglement $r$, with the value 1 or 0 for local entanglement or non-local entanglement.

This procedure looks much like the introduction of a spin 1/2 for atoms that were previously supposed spinless. Similarly, one uses here 2×2 matrices acting on local entanglement indices. Three matrices are of special significance, namely

$$P_0 = \begin{pmatrix} 0 & 0 \\ 0 & 1 \end{pmatrix}, P_1 = \begin{pmatrix} 1 & 0 \\ 0 & 0 \end{pmatrix}, S = \begin{pmatrix} 0 & 1 \\ 0 & 0 \end{pmatrix}. \qquad (3.2)$$

One sees that $P_1$ is a projection matrix, which is associated with local entanglement and the index 1, $P_0$ is also a projection, associated with non-local entanglement and index 0. The matrix $S$, which picks up the local entanglement index 0 and brings it to 1, represents a shift from non-local entanglement to local entanglement. One can also think of the analogy in Section 2, where local entanglement was associated with red color and non-local entanglement with white.

One will need no other matrix, and especially not the matrix $S^\dagger$, which is the adjoint of $S$ and would bring back local entanglement to non-local entanglement. This dismissal of $S^\dagger$ is characteristic of the idea of influence, which is itself a face of local entanglement: An influence of the particle $A$ on the gas can begin with a direct interaction of an atom with the particle $A$, or it can grow along a chain of transmission through successive interactions between atoms. It appears thus a quality that an atom can acquire and which can be transmitted to other atoms, but it can never be cancelled. When local entanglement is understood in that way, its growth is irreversibly transmitted from atom to atom, much like through a process of contagion. At a deeper level, one can see it also as a cluster property, as explained by Weinberg in Reference [20], book I, chapter 4.

One can express this idea formally by replacing the potential $U_{Aa}$, for interaction of the particle $A$ with an atom $a$, by the 2×2 matrix

$$\boldsymbol{U}_{Aa} = U_{Aa} \, (S_a + \, P_{1a}), \tag{3.3}$$

which brings non-local entanglement to local entanglement and also conserves it when it was previously present. Similarly, the two-body potential $V_{aa'}$ for interaction between two atoms $(a, a')$ can be replaced by a 4 ×4 local entanglement matrix:

$$\boldsymbol{V}_{aa'} = \, V_{aa'} \, (P_{0a} \otimes P_{0a'} + \, P_{1a} \otimes P_{1a'} + \, P_{1a} \otimes S_{a'} + S_a \otimes P_{1a'} \, ). \tag{3.4}$$

The first term in the parenthesis is concerned with two initially non-locally entangled atoms and keeps them non-locally entangled. The second one conserves similarly previously acquired local entanglement in the two atoms. The last two terms represent contagion, which occurs systematically when one locally entangled atom interacts with a non-locally entangled one and brings it to local entanglement.

When these operations are performed in all the two-body interactions in the composite system $AB$, the Hamiltonian $H$ in the Schrödinger equation (3.1) becomes an operator $H'$ involving now local entanglement in a way quite similar to what happens when the quantum dynamics of spinless atoms is extended to spin-1/2 atoms. Kinetic energy terms $-\nabla_a^2/2m$, which represent the free motion of atoms, conserve local entanglement and can be therefore considered as carrying simply a 2×2 unit matrix $I_a$.

This construction can be applied to any wave function and yields then a dynamics of local entanglement. Like in the analogy with spin, one does not deal any more with a unique wave function $\psi$ but with a set of many wave functions $\psi_s$, indexed by a string $s$ of $N$ local entanglement indices $(i_1, i_2, \ldots, i_N)$ taking either the values 0 or 1 (one denotes by $N$ the number of atoms in the gas). If one denotes more briefly the set $\{\psi_s\}$ by $\psi'$, its evolution is given formally by an equation of the type of Schrödinger's equation and is written as

$$i\hbar \partial \psi'/\partial t = H' \psi' , \tag{3.5}$$

where all these changes have been performed.

The operator $H'$ in (3.5) is now a $2^N \times 2^N$ matrix with elements consisting in differential operators and potentials $(U, V)$. A significant difference with usual circumstances is however that this evolution operator $H'$ is not self-adjoint. The absence of matrices $S_a^\dagger$, which would be associated with matrices $S_a$ in (3.4) and could bring back local entanglement to non-local entanglement, is responsible for this essential feature of $H'$.

Equation (3.5) makes sense nevertheless and it solutions exist, like they do for the standard Schrödinger equation (3.1). In a sense, Equation (3.5) draws only new consequences from the standard Schrödinger equation (3.1) by exhibiting how this equation can be used, after simple rewriting, to describe local entanglement in an explicit way.

The relation between the two evolution equations (3.1) and (3.5) goes farther and the standard Schrödinger equation (3.1) can also be derived from Equation (3.5) for the evolution of local entanglement. This inverted derivation is obtained by using the sum

$$\psi'' = \Sigma_s \, \psi_s, \qquad\qquad\qquad (3.6)$$

which is a function depending only on the positions of atoms and of the particle $A$, when all of them are spinless. One shows easily, by use of the explicit expressions (3.3-4) for interactions, that this function $\psi''$ satisfies the basic Schrödinger equation (3.1). Considering then initial conditions at a time zero, before interaction of the particle with the gas, one gets an equality $\psi''(0) = \psi(0)$, which is then obviously valid since all atoms are still non-locally entangled and local entanglement plays still no part in dynamics. At this time 0, the wave function $\psi_{s0}$ is the only non-vanishing one in the sum (3.6) (if one denotes by $s_0$ the index of the string where the local entanglement indices are (0000…0)). One has thus $\psi_{s0}(0) = \psi''(0) = \psi(0)$ at time $t = 0$ and, since the two functions $\psi(t)$ and $\psi''(t)$ obey the same Schrödinger equation (3.1), the equality $\psi(t) = \psi''(t)$ most be valid at all later times. *Equation (3.5), which describes the evolution of local entanglement, is therefore exactly equivalent to the standard Schrödinger equation* (3.1). The theory of local entanglement appears simply as a corollary of quantum dynamics, which it does not modify but to which it provides an extended interpretation.

One may also notice that symmetry properties in the components $\psi_s$ of the wave function $\psi$ result from the Bose-Einstein or the Fermi-Dirac symmetry between atoms. They are sometimes useful. One considers only the Bose-Einstein case to show these properties. A simple consideration of Equation (3.5) shows that every function $\psi_s$ is then symmetric under a permutation of two atoms ($a$, $a'$) carrying both index 1 in the string $s$, or under a permutation of two atoms carrying the same index 0. The symmetry between indices $a$ and $a'$ in the last two terms of the interaction (3.4) implies moreover that the final state of a collision, when it brings one of two colliding atoms to local entanglement, is symmetric between the atoms. This behavior is practically the same as in the argument by means of which Dirac [19] showed the impossibility of deciding which is which when two atoms have collided.

*Note*:

One adds a last useful remark to show how two-body collisions bring out local entanglement. In the case of a collision of the particle $A$ with an atom $a$ for instance, the Lipmann-Schwinger equation describing scattering [37] shows this result as follows. The collision involves an extension of the potential $U_{Aa} \rightarrow U_{Aa}$ as in (3.3) and brings out simply a replacement $T_{Aa} \rightarrow T_{Aa} = T_{Aa}(S_a + P_{1a})$ in the $T$-matrix (which is related to the $S$-matrix by $S = I + iT$ ). Scattering is therefore a privileged mode of transmission of local entanglement

*Irreversibility of local entanglement*

An essential feature of local entanglement is its irreversible character, which has several aspects: One of them was shown in the previous section by the irreversible waves of local entanglement, which one found to carry a contagion of local entanglement into the bulk of a gas. Without too much detail, one may consider obvious that all the atoms in the gas become locally entangled after a finite time, until total (algebraic) entanglement is reached in the full system, once and for all.

From a mathematical standpoint, this irreversible character is implied by the non-self-adjoint character of the evolution operator $H'$ in (3.5). One noticed that point already when mentioning that the first term in Equation (3.2), as well as the last two terms (3.3) in $H'$, involved the non-selfadjoint 2×2 matrix $S$ creating local entanglement, with no compensating terms that would have reestablished self-adjointness in $H'$. The evolution of local entanglement under Equation (3.5) is therefore absolutely irreversible, in view of the general relation between self-adjointness of the Hamiltonian and time reversibility of the Schrödinger equation.

One may notice also that this property of irreversibility is associated with the *history* of the two interacting systems, in analogy with Point 7 in Section 1. Moreover, if one is given the standard wave function $\psi$ at some time $t$, no operator can recover from it the components $\psi_s$ showing local entanglement of the various atoms. This point will turn out essential later on in this work and in its conclusions, because it means that local entanglement stands outside the domain of observables and relies on the full history of a wave function (or a density matrix) and not on any of its values at specific times.

*The field formalism for local entanglement*

A quantum field version of local entanglement can then be given for the sake of seeing its amount of generality. The existence of this field approach allows one to expect a wide range of validity for the notion of local entanglement, although its exact domain has not yet been thoroughly circumscribed [17].

One will only sketch the example of non-relativistic atoms obeying Bose-Einstein symmetry. Before introducing local entanglement, a convenient description consists in using creation and annihilation field operators, $\phi^\dagger(x)$ and $\phi(x)$, for of an atom, together with a quantum state $|\psi\rangle$ of the gas, given by [20]

$$| \psi \rangle = \int dx_1 \, dx_2 \, ... \, dx_N \, \psi(x_1, x_2, ... \, x_N) \; \phi^\dagger(x_1) \; \phi^\dagger(x_2)... \, \phi^\dagger(x_N) \, |0\rangle . \tag{5.7}$$

The wave function $\psi$ in the integral denotes the standard non-relativistic wave function of the system and every $x_j$ denotes the position of an atom. The reference state $|0\rangle$ is the vacuum state of quantum field theory, which satisfies the property $\phi(x)|0\rangle = 0$. The local fields $\phi$ and $\phi^\dagger$ satisfy the commutation relations

$$[\phi(x), \, \phi^\dagger(x')] = \delta(x - x'). \tag{3.8}$$

One can also introduce locally entangled quantum fields $\phi_r(x)$ with local entanglement indices $r = 0$ or 1. They satisfy the commutation relations

$$[\phi_r(x), \, \phi_r(x')] \, = 0,$$

$$[\phi_r(x), \, \phi_{r'}^\dagger(x')] = \delta_{rr'} P_r \, \delta(x - x'), \tag{3.9}$$

where $P_r$ is one of the matrices $P$ in (3.2). Another pair of field $(\alpha(x), \, \alpha^\dagger(x))$ can be used also to describe the particle $A$.

One can write down the previous evolution operator $H'$ showing the propagation of local entanglement as a field operator

$$H = H_{A0} + H_{B0} + V_{AB} + V_B. \tag{3.10}$$

The term $H_{A0}$ represents there the kinetic energy of the $A$-particle, $H_{B0}$ the kinetic energy of atoms, whether locally entangled or not. The term $V_{AB}$ describes the interaction of the particle $A$ with some atom, so that the outgoing state of this atom comes out locally entangled, notwithstanding whether it was initially locally entangled or not. The last term $V_B$ represents two-body interactions between atoms, with account of the contagious behavior of local entanglement as in (3.4).

When all the particles and atoms are non-relativistic and spinless, these operators are given by

$$H_{A0} = \int dy \; \alpha^\dagger(y)(-\nabla^2/(2m_A))\alpha(y), \tag{3.11a}$$

$$H_{B0} = \int dx \; \{\phi_1{}^\dagger(x)(-\nabla^2/(2m_a)) \; \phi_1(x) + \; \phi_0{}^\dagger(x)(-\nabla^2/(2m_a) \; \phi_0(x)\}, \tag{3.11b}$$

$$U_{AB} = \int dxdy \; \alpha^\dagger(y) \; \phi_1{}^\dagger(x)U(x,y)(\phi_1(x)+ \phi_0(x)) \; \alpha(y), \tag{3.11c}$$

$$V_B = (1/2) \int dxdx' \; \{\phi_0{}^\dagger(x) \; \phi_0{}^\dagger(x')V(x',x) \; \phi_0(x) \; \phi_0(x') + \phi_1{}^\dagger(x) \; \phi_1{}^\dagger(x')V(x',x) \; \phi_1(x) \; \phi_1(x')$$

$$+ \phi_1{}^\dagger(x) \; \phi_1{}^\dagger(x')V(x',x) \; \phi_1(x) \; \phi_0(x') + \phi_1{}^\dagger(x) \; \phi_1{}^\dagger(x')V(x',x)\phi_0(x) \; \phi_1(x')\}. \tag{3.11d}$$

Equation (3.10) can be extended easily to many types of particles, whether relativistic or non-relativistic, and also to many types of interaction. Although one does not consider this approach universal, one may expect it nonetheless to represent many kinds of interactions between a microscopic system and a macroscopic one. One will suppose that this generality is sufficient for expecting a wide range of validity for the notion of local entanglement in many practical cases.

One should also mention, to avoid misgivings, that the fields $(\phi_r(x), \phi_r{}^\dagger(x))$ are *not* operators in a definite Hilbert space (so that the self-adjoint quantities $\phi_r(x) + \phi_r{}^\dagger(x)$ or $\phi_r(x)\phi_r{}^\dagger(x)$ are not observables). A rigorous mathematical formulation of local entanglement would have presumably to make these fields act as operators within a sheaf of Hilbert spaces, in which every Hilbert space would represent quantum states of atoms with definite local entanglement. This extension towards the theory of sheaves has not been yet carefully attempted, so that the present approach remains in the mathematical framework of standard quantum physics and relies sometimes more on intuition that on proofs.

*Local probabilities of local entanglement*

One introduced earlier probabilities $f_1(x, t)$ and $f_0(x, t)$ for local entanglement and non-local entanglement of states of atoms near a point $x$. The purpose of the present subsection is mainly to anticipate the use that will be made of these quantities in later sections, and to show which mathematical meaning goes along with them when they are used in applications.

One will use for that purpose the initial state $|\psi\rangle$ of the macroscopic system before interaction as in Equation (3.7), except that non-locally entangled fields $\phi_0{}^\dagger(x)$ replace now the standard fields $\phi^\dagger(x)$. One lets then the operator $H'$ in (3.10-11) act on this initial state.

Because of the sums in equations (3.11c) and (3.11d) describing switches in local entanglement under interactions, or conservation, one gets a decomposition of a standard wave function $\psi_{AB}(t)$ for the composite $AB$ system as a sum upon locally entangled wave functions $\psi_\lambda(t)$ when the two systems interact.

Similarly, because a macroscopic system of actual interest is never in a pure state, it must be described generally by a density matrix $\rho_{AB}(t)$, which becomes a locally entangled matrix $\rho'_{AB}(t)$ under the previous construction of locally entangled wave functions. This kind of locally entangled state will be much used in the proposed theory of collapse. It evolves under the equation

$$i\hbar\partial\rho'_{AB}/\partial t = [H',\rho'_{AB}]. \tag{3.12}$$

The construction of local probabilities of local entanglement and of non-local entanglement proceeds then as follows. One introduces two operators, respectively for the total numbers of locally entangled and of non-locally entangled atoms, as

$$N_r = \int \phi^\dagger_r(x)\phi_r(x)dx, \tag{3.13}$$

with $r = 1$ or $0$. One can also introduce local operators for the number of locally entangled or non-locally entangled atoms $N_{r\beta}$ in a space cell $\beta$ in the gas, by means of the same integral as in (3.13) but with the points $x$ in that cell. The local probabilities for local entanglement and non-local entanglement, $f_1(x)$ and $f_0(x)$, where $x$ denotes now the center of a small cell $\beta$, are defined then as the ratios

$$f_r(x) = Tr\left(N_{r\beta}\rho'_{AB}\right)/N_\beta \quad, \tag{3.14}$$

where $N_\beta$ in the denominator denotes the average number of atoms in the cell $\beta$.

It should be stressed finally that the significant equation (2.4), which describes contagion, is still far from a rigorous proof. One will not elaborate however on that point, because it is partly due to the fact that the quantum statistical theory of irreversible processes, and particularly of this new type, stands itself on partially incomplete foundations. One will use nevertheless Equation (2.4) in the present work with no more argument than reliance on the satisfactory though often approximate success of similar considerations in many other domains, for practical purposes... The status of equations (2.1) and (2.2) is also pragmatic. One of the main difficulties in getting a rigorous theory of collapse would be to make all these constructions rigorous and one must acknowledge that the lack of these steps remains one of the main impediments in the present theory of collapse.

## 4. Fluctuations in environment and associated incoherence

The study of local entanglement was restricted in the previous sections to the case of a macroscopic system –a measuring device– and a microscopic measured system. The property of entanglement has no such restriction however [24, 3] and one will turn now to local entanglement between a measuring device and its environment

This study will provide the second building block of the present theory, in addition to local entanglement, and one will show that it consists in *the existence of a specific kind of incoherence, which arises in the quantum state of a macroscopic system under fluctuations in the action that it receives from its environment.*

### A model

One will deal with essentially the same model as used up to now. It consists again of a gas, made of atoms, enclosed in a solid box and mimicking a Geiger counter or a wire chamber. One will still denote this system by $B$ and consider now its environment, which is supposed to be a surrounding ordinary atmosphere, under standard conditions of temperature and pressure. No measurement is supposed to go on and one might say that this study deals with the state of a measuring $B$-system *before* a measurement.

A noticeable difference with the previous case is that now the molecules in the atmosphere stay out of the system and do not penetrate it to interact directly with the gas, like the particle $A$ did previously. This question can be considered in some detail but one will only sketch it: A collision of a unique molecule on the outside boundary of the box generates phonons in the solid wall, and these phonons carry local entanglement in the solid box while colliding together. They also bring out vibrations (surface phonons) on the inner face of the wall. These surface phonons, in their turn, perturb the boundary conditions for the wave functions of atoms in the gas and perturb accordingly these wave functions themselves, in the same way as one studied before.

Everything is a matter of collisions and transport in this example: Phonons behave essentially like particles and collide together in the solid box or along its boundary. Rather than speaking of perturbations of boundary conditions for the wave functions of atoms, one can consider just as well that there are collisions between surface phonons and atoms. The detailed physics of this simple device is therefore already rich and complex, like everything real, but the elementary effects occurring in it and on it are simply collisions, so that the collective behavior of the quantum state of $B$ can be legitimately envisioned as a rich combination of local entanglement effects, with multiple waves of local entanglement running inside it while remaining invisible, since no observable can manifest their presence.

### The status of environment

The particle $A$, which one saw interacting with the system $B$ in the previous sections, was considered to be a quantum system by itself. This description is no more obvious in the present case and a formal question, which is of importance, arises then namely: What is the theoretical status of the environment and can one consider it as a quantum system?

The quantum state $\rho_B$ of the gas was defined in Point 8 of Section 1 as a partial trace over a formal state $\left| \Psi_U \right\rangle \left\langle \Psi_U \right|$ of the universe and one can still use this definition. Although the evolution of this system $B$ cannot be separated from its interaction with environment, one may consider $B$ anyway as a well-defined quantum system, if only because its constitutive particles are themselves well defined and can be associated with a well-defined C*-algebra of observables. The Hilbert space of wave functions can be held then as a representation of this

algebra [38]. This is clear, at least conceptually, and one can still consider therefore the apparatus as a quantum system.

The theoretical status of environment is another matter. One can also define it in principle as a quantum system at a sharp time $t$. To do so, one considers it for instance to consist of all the molecules, present at time $t$ in some well-defined sphere $S$ enclosing the system $B$. A state $\rho_E(t)$ of this sharply defined environment at that sharp time is then again defined by a partial trace over $|\Psi_U\rangle\langle\Psi_U|$. The validity of this construction lasts however only a very short time. One cannot extend it over a sufficiently long time to allow a study and a use of waves of local entanglement in $B$. The molecules in the ideal sphere $S$ change many times during the motion of such a wave, some new molecules enter the sphere and other ones get out of it. The associated C*-algebra of observables changes also many times in an unpredictable way. One must therefore give up the prospect of identifying the environment with a quantum system when one deals with local entanglement.

*The case of a unique molecule*

One considers then for orientation the effect of a unique external molecule hitting the box and the local entanglement wave in the gas resulting from this collision. One found earlier that this wave of local entanglement has a finite velocity and spends a finite time in its motion, until entanglement becomes complete between the gas and the outgoing state of the molecule after collision. The previous discussion of waves of local entanglement shows that this time lapse is of order

$$\Delta t = L/c_s, \tag{4.1}$$

where $L$ denotes a typical linear size of the box and $c_s$ the velocity of the wave, which coincides in the gas with the velocity of sound.

This time delay will turn out essential in the present theory, even up to the existence of collapse. Presently, one considers it only as providing a time scale for the duration of an action of environment on the quantum state of $B$, which will be of great significance.

*The average action of environment*

Along with the previously defined instantaneous state $\rho_B(t)$ of the system $B$, one introduces now an average state $<\rho_B>$, which describes the macroscopic features of the system under the average action of environment. This procedure is familiar in statistical mechanics [39]. It consists in introducing a set of observables $\{A_k\}$, usually macroscopic ones, which describe the leading properties of a macroscopic system. One introduces also their average values $\{a_k\}$ as given data. If one considers them as relevant information for use in a theoretical study, a representative density matrix can be written as

$$\langle \rho_B \rangle = \exp\left\{-\sum_k \lambda_k A_k - \lambda_0 I\right\}, \tag{4.2}$$

where the values of the "Lagrange parameters" $\lambda_k$ are chosen to insure right values for the averages $\{a_k\}$, and $\lambda_0$ is introduced with a value insuring a unit trace for $<\rho_B>$

In the present example where the average action of the atmosphere reduces to the pressure that it exerts on the box, one has simply

$$\langle \rho_B \rangle = \exp\left\{-\left(H_B + H_{pB}\right)/kT - \lambda_0 I\right\}, \tag{4.3}$$

where $H_B$ is the Hamiltonian of the system $B$ and $H_{pB}$ expresses the average force that the atmosphere exerts on atoms in the outer face of the solid box.

One can use these expressions for averages to extract fluctuations out of $\rho_B$ by introducing the difference $\Delta\rho_B = \rho_B - <\rho_B>$, which contains these fluctuations. The trace of this matrix $\Delta\rho_B$ vanishes and one can separate its spectrum into positive and negative eigenvalues to get an expression

$$\Delta\rho_B = \rho_{B+} - \rho_{B-}, \tag{4.4}$$

where the matrices $\rho_{B+}$ and $\rho_{B-}$ are positive. This construction will play a central part in the theory and one writes down for that purpose at some time $t$ the expression

$$\rho_B(t) = <\rho_B(t)> + \rho_{B+}(t) - \rho_{B-}(t), \tag{4.5}$$

which will be used from there on.

*The nature of fluctuations*

The decomposition (4.5) of $\rho_B$ introduces positive fluctuations in $\rho_{B+}$ and negative ones in $-\rho_{B-}$. Their nature and their difference can be understood by considering their physical meaning. The action of environment in the present model consists in a flux of molecules hitting the box in various places at various times, and there are fluctuations in that flux. There are local end temporary slight excesses with respect to average, and also slight shortages, which are respectively at the origin of $\rho_{B+}$ and of $\rho_{B-}$.

One can also obtain estimates for some relevant parameters. The average number of collisions from external molecules on the box during a short time interval $\delta t$ is given by

$$\delta N = n_e v_e S \, \delta t, \tag{4.6}$$

where $n_e$ denotes the average number of atmospheric molecules per unit volume, $v_e$ their average velocity and $S$ the external area of the box (a factor $3^{-1/2}$ relating the average velocity of a molecule to its average component along a normal to the surface of the box is neglected). Values of $n_e$ around $10^{19}$-$10^{20}$ atoms per cubic centimeter, and of order $10^5$ cm per second for $v_e$, were often used in this work as representative.

The persistence of local entanglement effects, which is shown in (4.1) by their time of duration $\Delta t$, implies a permanent presence of many waves of local entanglement in the system $B$, with the large average number

$$N_W = n_e v_e S L / c_s, \tag{4.7}$$

The number of them entering into fluctuations is therefore

$$N_\phi = \Delta N_W = N_W^{1/2} = (n_e v_e S L / c_s)^{1/2}. \tag{4.8}$$

One can consider this quantity $N_\phi$ as the number of *fluctuating* waves of local entanglement that are present and moving at any time in the system $B$ (the index $\phi$ in its notation anticipates on a future connection with random phases).

The fluctuating nature of these waves of local entanglement yields useful information about the set of molecules that generated them. The number $N_W$ has no proper meaning as a number of waves, because any average must belong to the average matrix $<\rho_B>$ and the action of environment on it is expressed by Equation (4.3), which cannot be extended to local entanglement. As a matter of fact, there is no mathematical construction, either classical or of a quantum nature, and even no conceptual construction that would allow to distinguish and identify specific molecules, which generated the sample of fluctuating waves of local entanglement existing at some time $t$.

*Random phases in fluctuations*

The previous remark has significant consequences regarding a link between the fluctuating waves of local entanglement and an existence of incoherence in the state $\rho_B$ resulting from them.

The wave function of every molecule has a phase when hitting the box, so one looks at these phases for the molecules belonging to fluctuations. They turn out to be random. One can check this behavior by considering an academic example where the environment consists in a coherent beam of mono-energetic molecules. The phase of the wave function of every one of them, when it collides with the box, is perfectly defined by its place of arrival and its time of arrival. Nonetheless, the complete lack of a criterion distinguishing between average and fluctuations, and moreover between excesses and shortages in fluctuations, implies a random distribution of the place and time of arrival for molecules generating fluctuations and therefore random phases in their wave functions. This simple result, which remains certainly valid for more realistic types of environment, implies a general statement according to which *Fluctuations in the action of environment generate incoherence in the quantum state of a macroscopic system.*

*Locally entangled states*

A description of the quantum aspects of locally entangled states can be useful for better insight.

The external collisions producing this local entanglement as it stands at time $t$ must have occurred during the time interval $[t - \Delta t, t]$. Their number is therefore finite. If one denotes their set by $E$ (by which one means an "effective" part of the environment), there exists a density matrix $\rho_{BE}$, which is finite-dimensional since it involves a finite number of degrees of freedom in a bounded volume and a bounded energy.

One can thus introduce an "outgoing" locally entangled state $\rho'_B(t)$, which describes the state of the system $B$ together with outgoing states of all molecules that interacted with $B$ during the time interval $[t -\Delta t, t]$. Its eigenvectors exist and the history of local entanglement, as described in Section 3, allows to write down the associated wave functions as sums of locally entangled wave functions

$$\psi(s_1^{(1)}, s_2^{(1)}..., s_1^{(2)}, s_2^{(2)}, . ,...; ... ; x_1, x_2,...),\qquad(4.9)$$

which are analogous to the functions $\psi_s(x_1, x_2,...)$ in Section 3. They involve now $N$ variables of position $x_j$ and $N$ local entanglement indices $s_k^{(m)}$, where each one can take the values 0 or 1 (local entanglement or non-local entanglement) for each index $j$ specifying an atom. The upper index $m$ ranges in principle over all the molecules that collided with $B$ during the time interval $[t - \Delta t, t]$.

From Section 3, one derives that the eigenvectors of the standard density matrix $\rho_B(t)$ (showing not local entanglement) are linear combinations of the eigenvectors of $\rho'_B(t)$, which show local entanglement. The average locally entangled density matrix $<\rho'_B(t)>$ is trivial and coincides with $<\rho_B(t)>$, since $<\rho'_B(t)>$ is associated with fluctuations in the collisions that do not appear in $<\rho_B(t)>$.

All the effects of fluctuations, including local entanglement with environment and incoherence are therefore concentrated in the two matrices $\rho'_{B+}$ and $-\rho'_{B-}$.

One may also notice that these contributions from fluctuations do not extend generally to the whole content of the matrices $(\rho_{B+}, -\rho_{B-})$. These matrices are also present when the system $B$ is isolated but its state does not coincide at time 0 with the average $<\rho_B(0)>$ (which is also a time average in that case). One must be therefore careful in physical applications to evaluate which fraction of $\rho_{B+}$ and of $\rho_{B-}$ is actually associated with fluctuations, and not with an initial preparation of the system $B$. This is a difficult problem and as a matter of fact one which has not been solved explicitly in the present work.

The probability for incoherence of atomic states remains therefore a parameter, which one will denote by $W$ and on which little is known, except for an upper bound

$$W \leq 4/3\pi, \tag{4.10}$$

which is derived in the Appendix. It turns out unfortunately that the assumptions yielding the high value of $W$ in (4.10), or values close to this bound, are very exacting and leave us consider $W$ in the next sections as an unknown parameter, even in its order of magnitude.

## 5. Mechanism of collapse

One considers now the problem of collapse and use again, for definiteness, the model of $z$ measurement of a charged particle $A$ by a measuring counter $B$, which contains an atomic gas as its major active part. The particle is initially in a state of superposition (1.1) and the environment is an ordinary atmosphere.

The idea is that collapse would be a quantum phenomenon, depending only on quantum dynamics and resulting from it. Local entanglement would be the new tool renovating this study and it would act at two levels: Firstly in the growth of local entanglement between the two systems $(A, B)$ and, secondly, through generation of incoherence in the state of the macroscopic $B$ system. This incoherence, due to local entanglement of the system with fluctuations in its environment, would perturb the waves of local entanglement between $A$ and $B$ and generate fluctuations in the probabilities of measurement channels, with collapse as the end.

One might also say that this thesis assumes two mechanisms acting in collapse: an irreversible behavior of dynamics and an action of incoherence, the two of them having a unique origin in local entanglement.

To build up the relevant theory, formally, one will write down the density matrix for the composite system $A + B$ as a sum, like in Equation (4.5),

$$\rho'_{AB}(t) \ = <\rho'_{AB}(t)> + \ \rho'_{AB\,+}(t) - \ \rho'_{AB\,-}(t). \tag{5.1}$$

The "prime" index in these matrices means again that local entanglement is taken into account in them, so that their eigenfunctions are sums of wave functions showing local entanglement, as in Equation (3.6).

An "average matrix" $<\rho'_{AB}(t)>$ is introduced to serve as an indicator, by means of which the progress of collapse can be followed. One writes it down in the form

$$<\rho'_{AB}(t)> = \sum_j p_j(t) <\rho'_{Bj}(t)> \otimes |\,j\,\rangle\langle\,j\,|\ . \tag{5.2}$$

This expression is close to the description by standard measurement theory [1, 18, 21]. The matrix $<\rho'_{Bj}(t)>$ in (5.2) represents the average state of the measuring system, as it would stand at time $t$ if the measurement had started from the average state $<\rho_B(0)>$ involving the average macroscopic characters of $B$, and had interacted with the unique state $|\,j\,\rangle\langle\,j\,|$ of $A$, without intervention of the environment. The matrix (5.2) stands therefore as a reference to the measurement of a unique channel, except that one anticipates fluctuations in the quantum probabilities $p_j$ of various channels by introducing time-varying probabilities $p_j(t)$ in it. These probabilities are defined themselves in terms of the full density matrix at time $t$ as a trace

$$p_j(t) = Tr_B\langle\,j\,|\rho_{AB}(t)\,|\,j\,\rangle. \tag{5.3}$$

Finally, the prime in the notation $<\rho'_{Bj}(t)>$ means again that local entanglement with the system $A$ grew in $<\rho'_{AB}(t)>$, separately for each state $|\,j\,\rangle$, and generated probabilities of local entanglement $f_j(x, t)$ in it.

The two matrices $\rho'_{AB+}(t)$ and $\rho'_{AB-}(t)$ in (5.1) result then again from a separation of the eigenvectors in the difference $\rho'_{AB}(t) - <\rho'_{AB}(t)>$ into two sets, according to the positive or negative signs of their eigenvalues.

*Slips in coherence*

A key question regarding collapse asks how and why changes can occur in quantum probabilities of various measurement channels. Several attempts at understanding collapse were based on tentative answers to this question, with breaking of quantum laws, and the intent of the present approach is to show that such effects exist in quantum theory itself, without any essential modification in basic laws but taking local entanglement and environment into account. The basic phenomenon, which one will call a *slip in coherence*, occurs under the following conditions:

*Definition*: A collision between two atoms ($a, b$) is a slip in coherence when the state of Atom $a$ is locally entangled with some state $|\,j\,\rangle$ of the measured system and the state of Atom $b$ is non-locally entangled and incoherent.

The condition of incoherence for the second atom implies that its state belongs either to the matrix $\rho'_{AB+}(t)$ or -$\rho'_{AB-}(t)$. One will consider first the case where it belongs to $\rho'_{AB+}$. The state of the first atom $a$ must then also belong to $\rho'_{AB+}$ for an interaction between the two states to occur. Under these conditions, the significance of a slip in coherence stands in the following proposition:

*Proposition* When the slip in coherence occurs in the matrix $\rho'_{AB+}$, it generates transitions from the channels $j' \neq j$ to the channel $j$.

One will show more precisely that the average numbers of atoms $N_{j'}$, which are entangled with channels $j' \neq j$, decrease while the average number $N_j$ of atoms, entangled with channel $j$, increases correspondingly.

All these average numbers of entangled atoms are defined by the formula

$$N_k = Tr(\langle k \mid \rho_{AB} \int \phi^\dagger(x)\, \phi(x) dx \mid k \rangle), \tag{5.4}$$

which is used in the present case with the matrix $\rho'_{ABi+}$. If the collision occurs near a point $x$ in space, the probability for local entanglement of Atom $a$ with channel $j$ is $p_j f_j(x)$, whereas the probability for the state of Atom $b$ to be non-locally entangled is $f_0(x)$, and its probability for being incoherent (which carries a + sign in $\rho'_{ABi+}$) is $W$. Moreover, non-local entanglement of this initial state of $b$ implies that its probability for being entangled with any channel $k$ is $p_k$.

The probability that the collision in this slip occurs during a time interval $\delta t$ is $\delta t/2\tau$, where the factor $1/2$ accounts for a distinction between the two atoms $(a, b)$, which play different parts in the collision. This slip generates therefore changes in the average numbers of atoms, which are given by

$$\delta N_{j'} = -\, W p_{j'} p_j f_j(x)\, f_0(x)(\delta t/2\tau), \tag{5.5a}$$

$$\delta N_j = W p_j\, (1 - p_j) f_j(x)\, f_0(x)(\delta t/2\tau). \tag{5.5b}$$

Equation (5.5b) takes into account the fact that $b$ was already entangled with channel $j$ before the slip, and this is why the factor $1 - p_j$ enters in this expression. This is also the reason why Equations (5.5) bring no change in the total number of atoms.

The proposition regarding the effect of a slip holds therefore true in the case that one considered, as far as transitions in the numbers of entangled atoms are concerned.

## Collective effects of slips in coherence

One comes next to a point where the high complexity of incoherence plays a major part. One found in Section 4 that this incoherence comes from random phases, carried by moving waves of local entanglement, which arise themselves from fluctuations in the environment. One is dealing with them moreover as they occur in the matrix $\rho'_{AB+}(t)$, which results itself from an algebraic extraction of diagonal elements in the difference $\rho'_{AB}(t) - \langle \rho'_{AB}(t) \rangle$. One should remember that the number of these elements is of order $\exp(N)$ with $N$ the total number of atoms, each one of the corresponding eigenfunctions being a sum of about $\exp(N)$ locally entangled functions of the type that is shown in Equation (4.9).

The complexity of this construction allowing to state the problem is so high that the following assumption makes certainly sense: The distance of correlation between different space regions in the matrix $\rho'_{AB+}(t)$ cannot be much larger than a mean free path $\lambda$. This is an essential point in the discussion, which would need a rigorous proof for a full consistency of the proposed theory, but it looks obvious and one takes it for granted with no more analysis in the present work.

More precisely, one will rely on the following conjecture:

*Conjecture*: The restrictions $\rho'_{\beta+}(t)$ and $\rho'_{\beta'+}(t)$ under partial traces in distinct space cells $\beta$ and $\beta$ ', with size of order $\lambda$, of a matrix $\rho'_{AB+}(t)$ can be considered independent during a short time $\delta t$, with only small error.

As a matter of fact, an atom moves only over a distance $v\delta t << \lambda$ during the short time $\delta t$ that one is considering, and this conjecture is valid at a classical level. Its validity at a quantum level is more involved and will be attributed here mostly to the algebraic complexity of extracting eigenvectors with positive and negative eigenvalues when the matrices $\rho'_{AB+}$ and $\rho'_{AB-}$ are constructed.

This conjecture has two virtues. On one hand, it localizes the collective effects of slips within limited cells $\beta$: One can introduce for that purpose the local average numbers $N_{k\beta}$ of atoms in a cell $\boldsymbol{\beta}$, which are entangled with a state $|k\rangle$ of $A$, a space integral like in Equation (5.4) extending then over $\beta$. On the other hand, the quantities $N_{k\beta}$, which derive from quantum quantities with integer values, allow a use of the Poisson theorem for fluctuations $\Delta N = <N>^{1/2}$ of a positive integer $N$ with average $<N>$. One will also introduce for convenience formal localized probabilities of entanglement, defined by

$$p_{k\beta} = N_{k\beta} / N_{\beta}. \qquad (5.6)$$

Using these preliminaries, one can come back at the collective effects of all the slips in coherence, occurring in a cell $\boldsymbol{\beta}$ during a time $\delta t$. In the case of $\rho'_{\beta+}$, there are then two competing effects. In a first type, a channel $j'$ brings contributions to a channel $j$, and in a second type opposite transitions from the channel $j$ to the channel $j'$ occur. These two effects cancel on average, but the Poisson theorem implies simple expressions for standard deviations and correlations in the numbers of entangled atoms, of the type $(\Delta N)^2 = <N>$. One gets thus

$$\left\langle \left(\delta p_{j\beta}\right)^2 \right\rangle = p_j(1-p_j)W\int_\beta n_a f_j(x)f_0(x)dx\left(\delta t/\tau\right)/N_\beta^2, \qquad (5.7a)$$

$$\left\langle \delta p_{j\beta}\delta p_{j'\beta}\right\rangle = -p_j p_j W\int_\beta n_a\left(f_j(x)+f_{j'}(x)\right)f_0(x)dx\left(\delta t/\tau\right)/N_\beta^2, \text{ when } j\neq j' \qquad (5.7b)$$

A second kind of competition occurs between the effects belonging to $\rho'_{\beta+}$ and the ones in $-\rho'_{\beta+}$. Their combined effect is very simple and amounts only to a replacement of the factor $1/2\tau$ in Equations (5.7) by $1/\tau$.

When one takes finally into account independence of the events occurring in different cells $\beta$, one gets simple expressions of the standard deviations and the correlation coefficients for the fluctuations $\delta p_j$ of quantum probabilities in various channels, which are given by:

$$\langle(\delta p_j)^2\rangle = p_j (1 - p_j) \, W \, C_j (\delta t/\tau), \qquad (5.8a)$$

$$\langle\delta p_j \, \delta p_{j'}\rangle = - \, p_j \, p_{j'} \, W \, (C_j + C_{j'}) \, (\delta t/\tau). \qquad (5.8b)$$

The correlation coefficients $C_j$ in these expressions are given by

$$C_j = \sum_\beta \left(1/N_\beta^2\right) \int_\beta n_a f_j(x) f_0(x)\,dx\,, \qquad (5.9)$$

where

$$f_0(x) = 1 - \sum_j p_j f_j(x). \qquad (5.10)$$

One can get rough orders of magnitude regarding these expressions by introducing a typical size $L$ for the system $B$. Noticing that a well-developed wave of $AB$ local entanglement has a wave front with area of order $L^2$ and a width of order $\lambda$ where a product $f_j f_0$ is non-vanishing, the time scale of the process is of order $\tau L^2/n_a \lambda^5 W$. An upper bound on $W$ was shown to be $4/3\pi$ in the Appendix and a value $W = 0.1$ is eventually indicative, one can take for instance $n_a$ of order $10^{20} / cm^3$, $\lambda$ of order $10^{-5}$ $cm$ and $\tau$ of order $10^{-10}$ $s$. One thus gets a time scale for the collapse process of order $10^{-14}$ $s$ when $L = 1$ cm.

One will not pay much attention however to these estimates, in which enhancement of the signal (through ionization by an electric field for instance) is not taken into account. One will only retain as a conclusion that the effect under study is not negligible, in spite of a drastic simplification in its formulation.

*Transition from a time $t$ to a next one $t + \delta t$: The meaning of pointers*

Occurrence of fluctuations in the channel probabilities $\{p_j\}$ between times $t$ and $t + \delta t$ implies that the matrices $\rho_{AB+}$ and $\rho_{AB-}$ lose equality of their traces during this time lapse. In order to restart the process for a next step $\delta t$, one must define a new average matrix $<\rho_{AB}(t + \delta t)> .$ This is where the existence of a pointer enters significantly this approach.

Equation (4.2), which defined an average matrix $<\rho_B>$ can be used again to define a matrix $<\rho_{AB}(t + \delta t)>$. Various observables $\{A_k\}$ enter this construction, some of them as projection operators $|\, j\, \rangle\langle\, j\, |$ for the states of the measured system, with associated average values $p_j(t)$. Other significant observables $\{A_j\}$ in (4.2) denote "positions" of "pointers" by means of which the progress of measurement can be tested. The associated locally entangled matrix $<\rho'_{AB}(t + \delta t)>$ must also involve changes $\delta f_i(x,\ t)$ in the probabilities of local entanglement to get new matrices $\rho'_{AB\pm}(\ t + \delta t)$ resulting from an algebraic extraction of diagonal elements in the difference $\rho'_{AB}(t + \delta t) - <\rho'_{AB}(t + \delta t)>$.

A remarkable feature of this construction appears then: The presence of a pointer position among the objective observables $A_k$ defining average matrices makes one think of the decoherence effect, which depends on the positions of pointers and makes rapidly a matrix $\rho_{AB}(t)$ diagonal in the basis $|\, j\, \rangle$. This behavior is expected, but it stresses also a character of decoherence, already known but puzzling in the present case, which is that decoherence plays no part in the process of fluctuations in the quantum probabilities $\{p_j\}$. It would seem that decoherence and collapse both exist, but unrelated.

Their only common point is that both of them originate in actions of the environment, but these are quite different actions: There is an average one, very strong and yielding a rapid effect of decoherence [40], and there is a slower one, which originates in fluctuations of the environment and is expected to tend towards collapse.

This situation looks strange, and somewhat puzzling, at least when first encountered. Presently, one will only notice it without adding more comments.

*The stochastic mechanism of collapse*

The nature of collapse becomes clear when its effect is conceived as an accumulation of many successive fluctuations in the quantum probabilities $\{ p_j \}$. The randomness of these fluctuations, together with the linear behavior in $\delta t$ of the correlations in Equations (5.8), are characteristic of a stochastic process, which can be described by a Fokker-Planck equation

$$\partial \Phi / \partial t = \sum_{jj'} \partial_j \partial_{j'} \{ <\delta p_j \, \delta p_{j'}> \Phi \}, \qquad (5.11)$$

where the function $\Phi(p_1, \ p_2, \ldots; \ t \ )$ is a probability distribution, owing to incoherent random effects in the environment, of the quantities $\{p_j\}$ representing quantum "probabilities". In some sense, the notion of randomness has changed place. This distribution $\Phi$ starts at time 0, when the measurement begins, from

$$\Phi(0, \{ p_j \}) = \prod_j \delta(p_j - | c_j |^2). \qquad (5.12)$$

The boundary condition for Equation (5.11) is that $\Phi$ vanishes on the boundary of the $\{p_j\}$ domain, defined by $p_j \geq 0$, $\Sigma_j \, p_j = 1$.

*Boundary conditions and the problem of approach to collapse*

There is a difficulty, not yet fully mastered, in the theory at this stage. It comes from the dependence of the expression (5.8) for the correlation coefficients $\langle \delta p_j \delta p_{j'} \rangle$ on the quantities $\{ p_j \}$. They imply a Fokker-Planck current of probability with components

$$J_j = \partial_{j'} \{ <\delta p_j \, \delta p_{j'}> \Phi \} = \{ \partial_{j'} <\delta p_j \, \delta p_{j'}> \} \Phi + <\delta p_j \, \delta p_{j'}> \partial_{j'} \Phi. \qquad (5.13)$$

The difficulty arises when ones looks at this expression when $p_j$ tends towards zero. The first term in the right-hand side of (5.13) vanishes, because the condition $\Phi = 0$ for $p_j = 0$ is part of the boundary conditions for $\Phi$. The second term vanishes also in view of the expressions (5.8) for the correlation coefficients, which involve a factor $p_j$.

I must say that this difficulty was noticed only at a last moment in the writing of the present report. An answer came also, soon after (this kind of event occurred several times during the present research, in which many features are unfamiliar). Anyway, I guess that the answer stands in the fact that, ultimately, one is always dealing with quantities of the type $N_{k\beta}$ in (5.6), themselves associated with quantum quantities with a spectrum consisting in integers. One can then use the Poisson law, as it works for rare events. The standard deviation $\Delta N_{k\beta} = (<N_{k\beta}>)^{1/2}$ is much larger than the average $< N_{k\beta} >$ when this average is smaller than 1. The probability for vanishing of $N_{k\beta}$, and thus of the associated $p_{k\beta}$, is enforced rather than damped in that case and this is enough for making $p_k$ itself vanish. This means that a vanishing of a quantum probability is not governed by the Fokker-Planck equation (5.11), but by individual slips, or a few slips.

A detailed study of this random vanishing of a channel probability will not be worked out here, because one envisions it as a part in the framework of a more thorough study of collapse in realistic cases, which would be a next stage in the present theory. Presently, one may still hold it only as a conjecture, or a question mark in the theory, and proceed to a consideration of collapse as its consequence.

*Ultimate collapse*

Under this property of enforced disappearance for the probability of an individual channel, Pearle's theorem applies [11] (one may recall for illustration that this theorem extends widely a "gambler's ruin theorem" by Huygens according to which the probability for a gambler to win in a game of heads and tails is proportional to its staking, the other gambler becoming ruined). The probabilities for various channels are expected to vanish randomly during this process, one by one, until a unique one remains finally and, if it happens to be $p_k$, it reaches then the value $p_k = 1$ once and for all. There is then collapse on that channel.

In the present case, the stochastic probability for collapse in a specific channel $k$ is equal to the initial value of $p_k$ before measurement, *i.e.*, $| c_k |^2$ and this prediction coincides with the fundamental probability rule by Max Born.

One may notice that this randomness of collapse results from the random motion of the quantities $\{p_j(t)\}$, which is itself due to random fluctuations in the incoherent action of environment. One may still think nevertheless of Born's rule for microscopic events (such as collisions between atoms) because it brings out a legitimate logical representation of quantum processes, which extends and confirms an earlier one, derived from consistent histories [22, 33].

## 6. Additional comments

The main study in the present paper ends at this point, at least as far as matters of principle are concerned. Few numerical estimates have been made for getting actual predictions. They depend much on the actual value of the probability $W$ of incoherence but, anyway, are not much affected if this value is not much smaller than its upper bound $4/3\pi$.

The most interesting case is not the one that was considered here explicitly, with only waves of local entanglement moving in a gas. True physics is more encouraging here as a matter of fact than simplified models. The main difference, again in the case of a Geiger counter or a wire chamber, is the strong generation of ions and rapidly moving energetic electrons in a real detector. Among many differences with the present academic case, one is the restriction of local entanglement within the thin front of a wave in absence of an electric field, which must be compared with an extension of electrons and ions in a wide region where local entanglement grows almost everywhere at an approximately uniform rate. The effect is then much more efficient because of its extension and the process of collapse should be much more rapid.

A question, which links abstract inquiries with real measurements, may also be mentioned. It is concerned with experiments, like in a Stern-Gerlach device, where several detectors are used in a measurement. When elevated to the question of non-separablility in quantum physics, it refers to instruments with space-like separation when they measure, in the sense of relativity theory [41, 42]. There is no problem with these situations in the present approach and no trick but only an interpretation of the formalism of local entanglement. The space variable $x$, with which one dealt here, needs only be extended to the union of regions inside the various detectors, with local account of their difference, if any. Nothing else needs modification.

Another interesting question is concerned with the case of a decaying radioactive source and which asks more precisely: Why does one observe in that case a unique track, whereas a multitude of them is possible? I believe that this problem is related with the previous remark regarding cases where a channel probability is very small and can vanish easily. One might envision many possible tracks, all of them along straight lines [43] with very small probabilities, highly sensitive to disappearance. One would then be back rapidly to the case of a few channels (a few tracks) among which a unique one would win the game

One noticed earlier that, perhaps, collapse could occur in a macroscopic system (an inert gas for instance) with no macroscopic observable (no pointer) showing off a result and keeping memory of it. Were this behavior true, it would mean that many complete or partial collapse events occur in many places almost everywhere under innumerable circumstances. The existence of such effects could affect one's way of looking at Nature and at the global characters of quantum mechanics. Presently, one only mentions it.

The present lack of certainty regarding the exact rate of collapse raises another question: What would happen if collapse turned out to be a slow effect, at least in some circumstances? The answer from the present approach is clear, though maybe puzzling. Its essential point is that, in the present proposal, the events occurring in every measurement channel are absolutely unaffected by the process of evolution in the channel probabilities, until collapse. In every channel $j$, these events remain present at a macroscopic scale in the average density matrix $<\rho_{AB}>$ and they evolve exactly in the same way in channel $j$ as when the initial state of the measured system $A$ consists in the unique state $\left| j \right\rangle$.

One is back in some sense at the famous chain of measuring devices measuring other measuring devices, which was envisioned by Von Neumann [18]. The difference is that other systems around the first one, and from there other ones could participate in the growth of collapse until its completion. I dare say that this idea of temporary persistence of some events, among which perhaps beautiful or dreadful ones, until their waning forever has a romantic touch. But this is another matter.

One cannot omit as a last but one comment a relation of the proposed mechanism of collapse with Everett's conception of a wave function of the universe with multiple branches [7, 44]. Although the idea of a universal wave function was used for convenience in the premises in Section 1, it had no such consequence. There was no branching because, when a measurement occurs, there is some incoherence in the state of the measuring device, from its previous interactions with its own environment during the short though not extremely short time $\Delta t$ in (4. 1). A part of the universe at a distance larger than $c\Delta t$ cannot have therefore any influence on collapse. It can no more reach memory of the transient behavior of the collapsing process, because every such memory vanished everywhere with the vanishing quantum probability of its remembrance.

Finally, one cannot leave aside completely the meaning of collapse in the philosophy of quantum mechanics. The problem of collapse is by itself a philosophical problem regarding the consistency of quantum science. In spite of its significance, one will say that any comment of that kind of topics would be premature as long as the theory, which one would take as foundation, has not been questioned thoroughly from the technical standpoint of theoretical physics.

## 7. Conclusion

The initial intent of this work was to revisit the question of a possible self-consistency in quantum mechanics. Other attempts in that direction were made also in the recent years, with the proposal of Quantum Darwinism by Zurek and coworkers [45] and the Sub-ensemble Theory of Reduction by Allahverdyan, Balian and Nieuwenhuizen [46]. They are worth much attention but one will not compare them here with the present proposal and one will let this topic for later discussion.

The present work is definitely proposed as a conjecture, for reasons explained all along. I am far from pretending that the expected effects exist in the form in which they are proposed in this paper. Criticisms on the contrary are desirable or, still better, real proofs are needed! This form of collapse complements however so nicely previous hints, all of which arising from local entanglement alone, that the resulting construction, which came to mind as if building itself slowly, in its own way, is finally proposed here. Perhaps it will be shown an episode of fantasy among the wanderings of research, after examination. Perhaps also time will tell whether or not it belonged to science. It was anyway a long nice wandering amongst almost virgin lands.

### Acknowledgements

I thank Philippe Blanchard, Robert Dautray, Jürg Frölich, Phillip Pearle and Jean Petitot for their encouragement or help during this work, with special gratitude to the memory of Bernard d'Espagnat. Assistance by Franck Laloë in clarifying these ideas was particularly precious. I also benefitted from lucid criticisms, from Stephen Adler, Heinz-Dieter Zeh and Phillip Pearle, on earlier proposals of mine on the origin of incoherence.

### Appendix

### A bound on the probability of incoherence $W$

This appendix is concerned with the probability $W$ for incoherence. One will rely and its average density matrix $<\rho_B>$ has the simple form (3.2). The results in Section 3 show that a high disorder results from the action of environment (at least when the product $n_e v_e$ in (3.6) is not very small). The eigenvectors of $\rho_B$ at a sharp time are then sums of many incoherent terms with correlation ranges of order a mean free path. One will say that this situation is one of "high disorder", which makes the density matrix $\rho_B$ random, "for all practical purposes". Another feature making this randomness much stronger is the very high degeneracy of the eigenvectors of $<\rho_B>$, which makes them extremely sensitive to even very small perturbations.

One may assume however, in the simple model under consideration, that the energy distribution in (3.2) is not appreciably affected by fluctuations in the environment so that thermal equilibrium is not appreciably disturbed.

As a first step in the study of $\rho_B$, one uses this stability in the energy distribution by splitting $r_B$ into a direct sum of terms $\rho_{Bn}$, associated with separate energy intervals $[E_n, E_n + \Delta E]$ with $\Delta E$, not too large. One has then simply

$$\rho_B = \prod_n \rho_{Bn}. \tag{A.1}$$

One will denote by $N$ the dimension of this finite matrix. This number is very large in spite of the smallness of $\Delta E$, because of a very high degeneracy in the energy spectrum.

One will denote also by $|\mu\rangle$ the eigenvectors of $\rho_n$, and by $p_\mu$ the corresponding eigenvalues. Similarly, one denotes by $<\rho_n>$ the matrix resulting in the same way from an average $<\rho_{Bn}>$ in $<\rho_B>$ and by $|m\rangle$ its eigenvectors (which are also eigenvectors of the Hamiltonian $H_B$). All the eigenvalues of $<\rho_n>$ have then the same value $p = 1/N$.

High disorder implies a strong random behavior of $\rho_n$ with two main consequences: (*i*) The orientation of the orthogonal basis of vectors $\{|\mu\rangle\}$ with respect to the basis $\{|m\rangle\}$ is random. (*ii*) The distribution of the positive eigenvalues $\{p_\mu\}$ is random with average value $p$ (this value resulting from the unit trace of $\rho_n$).

One will also make a last assumption, which expresses a complete randomness in the distribution of eigenvalues $p_\mu$, and resulting again from the extreme instability in the quantum states of $<\rho_n>$. This strong assumption is the following assumption:

***Assumption of maximal randomness***: The probability distribution for the random values $p'$ of the eigenvalues $\{p_\mu\}$ is given by

$$g(p') = (1/p)\exp(-p'/p). \qquad (A.2)$$

One will consider that this assumption is justified by a complete lack of determination in the individual eigenvectors $|\mu\rangle$, which implies a minimal value for the corresponding measure of information (algorithmic entropy)

$$-\int g(p') \log[g(p')] \, dp'. \qquad (A.3)$$

When writing the equation expressing a minimum for (A.3) and taking account of the average value $p$ for the random variable $p'$, one gets (A.2). One may notice also that this result implies the following value for the standard deviation $\Delta p'$:

$$(\Delta p')^2 = p'. \qquad (A.4)$$

One can then state the main consequence of these preliminaries. which is expressed by the following lemma

***Lemma***: When written as an $N \times N$ matrix in the basis $\{|m\rangle\}$ of eigenvectors of the Hamiltonian $H_B$, the matrix $\Omega = \rho - <\rho>$ is a normalized Wigner random matrix [47].

One recalls that a self-adjoint matrix $\Omega$ with dimension $N$ is a (normalized) Wigner random matrix when all its matrix elements are independent complex random numbers $\omega_{mm'} = \langle m|\Omega|m'\rangle$, real for $m = m'$. If one denotes by $Av$ the operation of taking average values on functions depending on these numbers, one gets the relations

$$Av(\omega_{mm'}) = 0, \qquad Av(|\omega_{mm'}|^2) = 1/N, \qquad (A.5)$$

where the normalization $1/N$ in the second equation results form normalization. More general Wigner matrices correspond to cases where the right-hand side is a constant though not necessarily equal to $1/N$.

The supposed properties of randomness in the eigenvalues of $\rho_n$ and in the orientation of its eigenvectors imply that the difference $\Delta\rho_n = \rho_n - <\rho_n>$ should be a Wigner random matrix. There is however a constraint, which is due the exactly zero trace with no fluctuation of $\Delta\rho_n$. This condition is not satisfied by a strictly defined Wigner matrix, whose trace fluctuates. The relevant corrections are however of order $1/N$ on every prediction that one will be using and is therefore negligible: One recalls that the number $N$ of states under consideration is an exponential in the number of atoms in the system and that the normalization $1/N$ results from the property $Tr(\Delta\rho_n^2) = (\Delta p)^2 = p = 1/N$.

If one writes down then the eigenvalues $q'$ of $\Delta\rho_n$ in the form $q' = px$, Wigner's "semicircle theorem" [47] determines the probability distribution $h(x)$ for the values of $x$, with $-2 \leq x \leq 2$, which is given by

$$h(x)dx = (4 - x^2)^{1/2} \, dx/2\pi. \tag{A.6}$$

This distribution implies an average value $4/3\pi$ for the positive values of $x$ and the opposite for the negative values, with From there on, one can go back to from this result on $\rho_{Bn}$ to the matrix $\rho_B$ and conclude, after separating the positive and negative parts in the difference $\rho_B - <\rho_B> = \rho_{B+} - \rho_{B-}$, the simple result

$$W = Tr(\rho_{B+}) = Tr(\rho_{B-}) = 4/3\pi. \tag{A.7}$$

The assumptions, which lead to this result, must be taken with caution however and this is why one considers them only as an extreme case and an upper bound.